\documentclass{aa}  
\usepackage{graphicx}
\usepackage{txfonts}

\newcommand\Swift{{\it Swift}}

\newcommand\kms{\ifmmode {\rm km\ s}^{-1} \else km s$^{-1}$\fi}
\newcommand\Hubble{\ifmmode {\rm km\ s}^{-1}\ {\rm Mpc}^{-1}
    \else km s$^{-1}$ Mpc$^{-1}$\fi}
\newcommand\ctssec{\ifmmode {\rm cts\ s}^{-1} \else cts s$^{-1}$\fi}
\newcommand\ergsec{\ifmmode {\rm ergs\ s}^{-1} \else
    ergs s$^{-1}$\fi}
\newcommand\eflux{\ifmmode {\rm ergs\ s}^{-1}\;{\rm cm}^{-2} \else
    ergs s$^{-1}$ cm$^{-2}$\fi}
\newcommand\phflux{\ifmmode {\rm photons\ s}^{-1}\;{\rm cm}^{-2}
    \else   photons s$^{-1}$ cm$^{-2}$\fi}
\newcommand\efluxA{\ifmmode {\rm ergs\ s}^{-1}\;{\rm cm}^{-2}\;{\rm
    \AA}^{-1} \else ergs s$^{-1}$ cm$^{-2}$ \AA$^{-1}$\fi}
\newcommand\efluxHz{\ifmmode {\rm ergs\ s}^{-1}\;{\rm cm}^{-2}\;{\rm
    Hz}^{-1} \else ergs s$^{-1}$ cm$^{-2}$ Hz$^{-1}$\fi}
\newcommand\cc{\ifmmode {\rm cm}^{-3} \else cm$^{-3}$\fi}
\newcommand\FWHM{\ifmmode {\rm FWHM} \else ${\rm FWHM}$\fi}
\newcommand\Msun{\ifmmode M_{\odot} \else $M_{\odot}$\fi}
\newcommand\Lsun{\ifmmode L_{\odot} \else $L_{\odot}$\fi}

\newcommand\Hbeta{\ifmmode {\rm H}\beta \else H$\beta$\fi}
\newcommand\Kalpha{\ifmmode {\rm K}\alpha \else K$\alpha$\fi}
\newcommand\NH{\ifmmode N_{\rm H} \else N$_{\rm H}$\fi}

\begin{document}
   \title{
   The multiwavelength afterglow of GRB~050721:
   a puzzling rebrightening seen in the optical but 
   not in the X-ray.
   \footnote{Based on observations acquired with ESO telescopes at Paranal, 
   under programs 275.D-5022 and 075.D-0787}}


 \author{
L. Angelo Antonelli \inst{1,2},
Vincenzo Testa\inst{1},
Patrizia Romano\inst{3},
Dafne Guetta \inst{1},
Ken'ichi Torii \inst{4},
Valerio D' Elia \inst{1},
Daniele Malesani \inst{5},
Guido Chincarini\inst{3,6},
Stefano Covino\inst{3},
Paolo D'Avanzo\inst{3},
Massimo Della Valle\inst{7,8},
Fabrizio Fiore \inst{1},
Dino Fugazza \inst{3},
Alberto Moretti\inst{3},
Luigi Stella\inst{1},
Gianpiero Tagliaferri\inst{3},
Scott Barthelmy\inst{9},
David Burrows\inst{10},
Sergio Campana\inst{3},
Milvia Capalbi\inst{2},
Giancarlo Cusumano\inst{11},
Neil Gehrels\inst{9}, 
Paolo Giommi\inst{2},
Davide Lazzati \inst{12},
Valentina La Parola\inst{11},
Vanessa Mangano\inst{11},
Teresa Mineo\inst{11},
John Nousek\inst{10},
Paul T. O'Brien\inst{13}, 
Matteo Perri\inst{2}.
           }
            
\offprints{L. Angelo Antonelli -- \email{a.antonelli@oa-roma.inaf.it}}

\institute{
 INAF-Astronomical Observatory of Rome, via Frascati, 33, I-00040, Monte Porzio Catone (Rome), Italy
\and
ASI Science Data Center, Via Galileo Galilei, I-00044 Frascati (Rome), Italy
\and
INAF-Astronomical Observatory of Brera, via E. Bianchi, 46, Merate (LC), I-23807, Italy
\and
Department of Earth and Space Science, Graduate School of Science, Osaka University,\\
1-1 Machikaneyama-cho, Toyonaka, Osaka 560-0043, Japan
\and
International School for Advanced Studies (SISSA-ISAS), Via Beirut 2-4, I-34014 Trieste, Italy
\and
Universit\`a degli Studi di Milano ``Bicocca", Piazza delle Scienze, 3, I-20126, Milano, Italy
\and
INAF-Astrophysical Observatory of Arcetri, Largo E. Fermi, 5,  I-50125 Firenze, Italy
\and
Kavli Institute for Theoretical Physics University of California Santa Barbara, CA   93106
\and
NASA/Goddard Space Flight Center, Greenbelt, MD 20771
\and
 Department of Astronomy \& Astrophysics, 525 Davey Lab., Pennsylvania State University, University  Park, PA 16802, USA
\and
INAF Ð Istituto di Astrofisica Spaziale e Fisica Cosmica Sezione di Palermo, Via Ugo La Malfa 153, I-90146 Palermo, Italy
\and
JILA, University of Colorado, Boulder, CO 80309, USA
\and
Department of Physics and Astronomy, University of Leicester, Leicester LE 1 7RH, UK
 }

   \date{Received --- ; accepted --- }

      
 \abstract
   {GRB\,050721 was detected by \Swift\ and promptly followed-up, in the X-ray by Swift itself 
   and, in the optical band, by the VLT operated, for the first time, in rapid response mode 
  starting observations about 25 m after the burst. 
   A multiwavelength monitoring campaign was performed in order to study its afterglow's 
   behavior.}
   {We present the analysis of the early and late afterglow emission in both the X-ray and optical 
   bands, as observed by $\Swift$, a robotic telescope, and the VLT. We compare early observations
   with late afterglow observations obtained with \Swift\ and the VLT in different bands in order
   to constrain the density of the medium in which the fireball is expanding.}
   {We have analyzed both the X-ray and the optical light curves and compared the
   spectral energy distribution of the afterglow at two different epochs. }
   {We observed an intense rebrightening in the optical band at about one day after
   the burst which was not seen in the X-ray band. This is the first observation of a GRB
   afterglow in which a rebrightening is observed in the optical but not in the X-ray band. 
   The lack of detection in X-ray of such a strong rebrightening at lower energies 
   may be described with a variable external density profile. In such a scenario, the 
   combined X-ray and optical observations allow us to derive that matter located at 
   $\sim10^{17}$ cm from the burst is about a factor of 10 higher than in the inner region.
   }
   {}
 
  
  \titlerunning{The puzzling case of the GRB 050721 afterglow}
\authorrunning{L.A. Antonelli et al.}

  \maketitle
%

\section{Introduction}

The \Swift\ Gamma-ray Burst Explorer (\cite{gehr04}) is currently detecting
2--3 gamma-ray bursts (GRBs) per week, distributing coordinates with
very small uncertainties (few arcmin down to several arcsec) with delays
ranging from few seconds to tens of seconds after the GRB event. Thanks to
its fast-pointing capabilities, \Swift\ is able to perform
observations of the GRB early afterglow phases, both in the X-ray and
ultraviolet/optical bands. Moreover, the prompt \Swift\ alerts also allow
follow-up of GRBs with ground-based facilities. In particular the
European Southern Observatory (ESO) made its four Very Large
Telescope (VLT) units able to quickly react to transient sources,
allowing them to repoint and start observations within just 8 min 
after the trigger, by developing the Rapid Response Mode (RRM) procedure.
Before \Swift\ the time needed to determine the GRB position was large 
and most afterglow measurements could start hours after the burst. Thanks 
to \Swift\ we can now investigate the characteristics 
of the very early stages of the afterglow,  when the physical properties of 
the fireball and of the circumburst medium can be derived from the 
properties of the light curve. Moreover, early observations easily span a long 
dynamical range in the afterglow lifetime, so that a rich phenomenology 
can be observed.

Here we report our analysis of GRB\,050721, a long, weak burst
discovered and located by the Burst Alert Telescope 
(BAT; \cite{barthelmy05}) on board $\Swift$ and immediately (186~s 
after the burst) followed--up with the $\Swift$  narrow field instruments: 
the X-ray Telescope (XRT; \cite{burr05}) and the UV Optical Telescope 
(UVOT, \cite{roming05}). A previously unknown fading X-ray source
was detected within the BAT error circle, while no evident optical
counterparts were seen in the optical images obtained by UVOT
(\cite{anto05}). The fast accurate localization allowed a 0.3~m
robotic telescope, located in New Mexico, to detect the optical
afterglow (OA) $\sim 369$~s after the trigger (\cite{tori05}),
at the level of  $I_c \sim 15.6$ mag. The field was also observed 
by the {\it MISTICI} collaboration with the ESO-VLT UT2 telescope 
operated for the first time in rapid response mode (RRM).  The VLT
observations started 25 min after the GRB and confirmed the presence
of the OA within the XRT error circle (\cite{covi05}).
Such a prompt identification of the OA allowed a very dense sampling of its
light curve at early times, making it one of best examples ever obtained
(\cite{test05}). A spectroscopic observation was also performed at VLT but the
OA spectrum was heavily contaminated by the contribution of a bright
($R = 16.7$ mag) foreground star lying very close (1.4\arcsec) to the OA. No
useful information could be extracted from it. 

The OA was extensively observed for several days after the burst with both \Swift\ 
and the VLT, obtaining a very good multiwavelength coverage. 

In the next sections a detailed study of the XRT 
and the VLT data is presented and discussed.

\section{GRB\,050721}

GRB\,050721 was detected by BAT on 2005, July 21 at 04:29:14.3 UT and
the BAT on-board calculated location was ${\rm RA} = 16^{\rm h} 53^{\rm m}
47^{\rm s}$, ${\rm Dec} = -28\degr 23\arcmin 22\arcsec$ (J2000), with an
uncertainty of 3\arcmin{}  error radius (90\% containment, \cite{anto05}). 
The refined BAT ground
position was ${\rm RA} = 16^{\rm h} 53^{\rm m} 48\fs5$, ${\rm Dec} =
-28\degr 23\arcmin 10\arcsec$ (J2000), with an error radius of 3\arcmin{}
(90\% containment, statistical and systematic). The partial coding
was 13\% (\cite{feni05}).

The masked-weighted light curve shows a FRED-like structure with a
single large peak starting to rise at $T_0-5$~s
($T_0$ being the trigger time), peaking at $T_0+3.7$~s
and decaying back to background levels by $T_0 +50$~s. The
peak is visible in the 15--100 keV energy band, but not at higher
energies. The calculated $T_{90}$ (15--350 keV) is ($39 \pm 2$)~s
(estimated error including systematics).

The power-law photon index of the time-averaged spectrum was
$1.81 \pm 0.08$. The fluence in the 15--350 keV band was
$(5.05\pm0.25) \times 10^{-6}$~erg~cm$^{-2}$. The 1-s peak photon flux
measured from $T_0+3.7$~s in the 15--350 keV band was
$(3.4\pm0.8)$~ph~cm$^{-2}$~s$^{-1}$. All the errors are quoted at the
90\% confidence level (\cite{feni05}).

        \begin{figure}
         \centering
         \includegraphics[width=9cm]{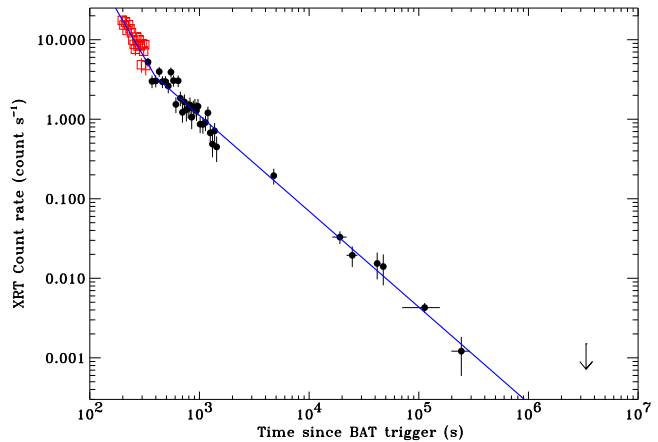}
                \caption{X-ray light curve of GRB\,050721 afterglow in the 0.2--10~keV energy band.
	Empty squares and filled circles indicate WT and PC data, respectively.
         The curve is background-subtracted and the time is referred to the BAT trigger,
         2005 Jul 21 at 04:29:14.28 UT. The solid line shows the best-fit broken power law (see Sect.~3.2).
         The last point is a 3-$\sigma$ upper limit.
                }
                \label{grb050721:xrtlcv}
        \end{figure}

\begin{table*}[htbp]
\centering
\caption{Log of Swift XRT observations.}
\begin{tabular}{cccccc}
\hline
Sequence & Obs. mode    & Start time          & End time            & Exposure  & Time since trigger \\ \hline
         &              & yyyy-mm-dd hh:mm:ss & yyyy-mm-dd hh:mm:ss & s         & s             \\
\hline
00146970000 &   XRT/IM  & 2005-07-21 04:32:20 & 2005-07-21 04:32:22 & 2.5       &     186       \\
00146970000 &   XRT/WT  & 2005-07-21 04:32:28 & 2005-07-21 05:48:30 & 205       &     194       \\
00146970000 &   XRT/PC  & 2005-07-21 04:34:38 & 2005-07-21 05:50:49 & 1291      &     324       \\
00146970001 &   XRT/PC  & 2005-07-21 09:00:03 & 2005-07-21 19:01:44 & 8818      &   16249       \\
00146970003 &   XRT/PC  & 2005-07-22 00:05:07 & 2005-07-22 23:42:59 & 19169     &   70554       \\
00146970004 &   XRT/PC  & 2005-07-23 11:27:27 & 2005-07-24 13:13:58 & 5937      &  197894       \\
00146970005 &   XRT/PC  & 2005-08-28 14:50:42 & 2005-08-29 08:52:59 & 8600      & 3320489       \\
\hline
 \end{tabular}
 \label{grb050721:obs}
 \end{table*}

\section{XRT data analysis}

The XRT observations of GRB\,050721 started on July 21, 2005 
at 04:32:20 UT, only 186~s after the trigger, and ended on July 28
at 08:52:59 UT,
thus summing up a total net exposure (in photon counting (PC)
mode) of  $\sim 35\,000$~s spread over a $\sim 4$~d baseline
(see Fig.~\ref{grb050721:xrtlcv}).
The monitoring was organized in 7 observations.
The first one was performed as an automated
target (AT) with XRT in auto state. Therefore, during
this observation the automated mode switching made XRT take an
initial 2.5~s image (IM at $T_0+186$~s) followed by ($T_0+194$~s) a series of
windowed timing (WT) frames
which were taken until the on-board measured count rate was low enough
for XRT to switch to PC mode ($T_0+324$~s). After this, XRT remained
in PC mode. The log of Swift observations used for this work is
summarized in table~\ref{grb050721:obs}.

The XRT data were processed with
the task {\tt xrtpipeline} (v0.9.9), applying standard calibration and filtering 
and screening criteria, i.e., we cut out temporal intervals during which the CCD
temperature was higher than $-47$~$^\circ$C, and we removed hot and
flickering pixels.
An on-board event threshold of $\sim$0.2~keV was also applied to the central pixel,
which has been proven to reduce most of the background due to either
the bright Earth limb or the CCD dark current (which depends on the
CCD temperature). For our analysis we further selected XRT grades
0--12 and 0--2 for PC and WT data, respectively (according to Swift
nomenclature).

\subsection{Spatial analysis}

An uncatalogued, fading X-ray source was detected within the BAT error circle.
We used the {\tt xrtcentroid} task (v0.2.7) to derive the afterglow position, obtaining ${\rm RA} 
=16^{\rm h} 53^{\rm m} 44\fs62$, ${\rm Dec}=-28^{\circ} 22\arcmin 52\farcs1$ (J2000). 
We estimate the uncertainty to be 3\farcs3 (90\% confidence level). This position takes 
into account the correction for the misalignment between the telescope and the satellite 
optical axis (\cite{centroids}).

\subsection{Temporal analysis}

During the first orbit of the XRT observation the intensity of the
afterglow was high enough to cause pile-up in the PC mode data.
To account for this effect we extracted the source events in an annulus
with a 30-pixel outer radius ($\sim71\arcsec$) and a 4-pixel inner
radius. For the PC data collected after the first orbit, the entire circular
region (30 pixel radius) was used, instead.  The WT data were
extracted in a rectangular region $40\times20$ pixels along the 
image strip. The selected extraction regions correspond to
$\sim 43$\% (piled-up PC), $\sim 95$\% (non piled-up PC), and
$\sim 95$\% (WT) of the XRT PSF. To account for the background,
data were also extracted in PC mode within an annular region (inner 
and outer radii of 50 and 100 pixels) centered on the source, and in 
WT mode within a rectangular box ($40 \times 20$ pixels) far from 
background sources. Figure~\ref{grb050721:xrtlcv} shows the 
background-subtracted light curve extracted in the 0.2--10 keV 
energy band, with the BAT trigger as origin of time. The last point 
is a 3-$\sigma$ upper limit. 
A fit with a broken power law $F(t)=K t^{-\alpha_1}$  for $t<t_{\rm b}$ and
$F(t) = K t_{\rm b}^{-\alpha_1} \, (t/t_{\rm b})^{-\alpha_2}$ for $t>t_{\rm b}$, 
where $t_{\rm b}$ is the time of the break, yields slopes $\alpha_1=2.37\pm0.24$ and $\alpha_2=(1.20\pm0.04)$,
and a break at $t=399_{-35}^{+75}$ s after the BAT trigger 
($\chi^2_{\rm red}=1.39$; 52 d.o.f.).  The late afterglow (i.e the last 3 
observations reported in Table~\ref{grb050721:obs}) light curve is fitted 
by a simple power law $F(t) = K t^{-\alpha}$ with $\alpha=1.20^{+0.39}_{-0.42}$ .


        \begin{figure}
        \centering
        \includegraphics[width=5.5cm, angle=270]{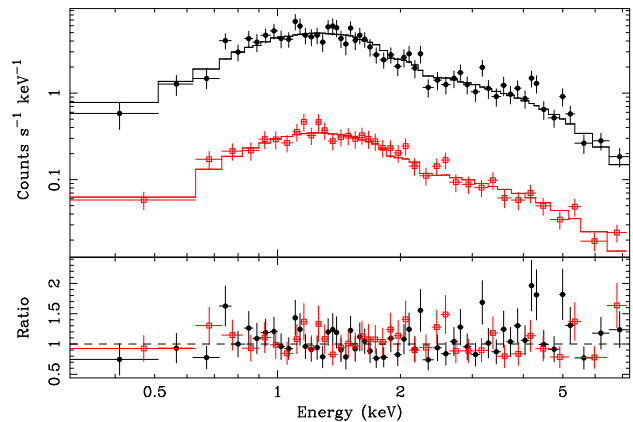}
        \caption{X-ray spectrum of the afterglow of GRB\,050721.
        {\bf Top:} spectra of WT (filled circles)
        and PC data (empty squares) fitted with an absorbed power law model.
        {\bf Bottom:} residuals from the simultaneous power-law fit to
        all the data.
         }
        \label{grb050721:xrtspec}
        \end{figure}


        \begin{figure}
    \centering
          \includegraphics[width=6cm, angle=270]{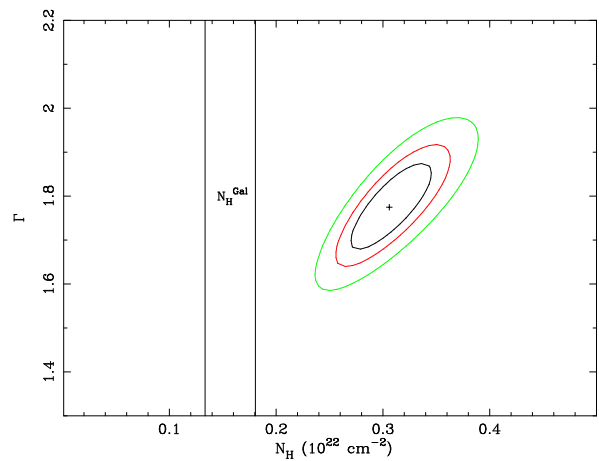}
          \caption{Contour plot of the column density vs.\ the power law photon index for the
        X-ray spectrum of GRB 050721. The contours refer to 1-, 2- and 3-$\sigma$ confidence levels.
        The Galactic column density ( $1.60 \times 10^{21}$~cm$^{-2}$; \cite{kalberla05}) is reported
        with an uncertainty of 15\% (solid vertical lines). }
         \label{grb050721:xrtcont}
        \end{figure}

\subsection{Spectral analysis}

Spectra of the source and background were extracted in both the
WT mode and PC mode in the same regions described above at the 
time corresponding to the first orbit. Ancillary response files were 
generated with the task {\tt xrtmkarf} within FTOOLS v6.0.4, and 
account for different extraction regions and PSF corrections.  
We used the latest spectral redistribution matrices (RMF, v008).  
The 0.3--10 keV band spectra were rebinned with a minimum of 
20 counts per energy bin to allow the use of $\chi^2$ and fitted by adopting 
a simple model of an absorbed power law model, with the Hydrogen 
column density (at $z=0$) kept as a free parameter. In the early part of the observation
(first 324 s) the WT mode spectrum yields a photon index 
$\Gamma=1.71_{-0.13}^{+0.14}$ and a column density 
$N_{\rm H}=2.92_{-0.49}^{+0.55} \times 10^{21}$ cm$^{-2}$
(slightly higher than the Galactic value: $1.57 \times 10^{21}$~cm$^{-2}$;
\cite{dicklock90}; $1.60 \times 10^{21}$~cm$^{-2}$: \cite{kalberla05}),
$\chi^2_{\rm red}=1.00$; 55 d.o.f. The observed count rate was 
($10.0\pm0.3$) cts s$^{-1}$ corresponding to an unabsorbed flux 
$F_{\rm X}$(0.2--10 keV)=$7.2\times10^{-10} \eflux$.  
In the remaining part of the first orbit (324--1450~s) the PC mode spectrum yelds
a photon index $\Gamma=1.86_{-0.16}^{+0.17}$ and a column density 
$N_{\rm H}=3.25_{-0.62}^{+0.74} \times 10^{21}$ cm$^{-2}$,
$\chi^2_{\rm red}=0.67$; 36 d.o.f. The observed count rate was
($0.72\pm0.02$) cts s$^{-1}$ corresponding to an unabsorbed flux of 
$F_{\rm X}$(0.2--10 keV)=$1.40\times10^{-10} \eflux$.
In order to enhance statistics, we also performed a simultaneous fit of 
the WT data and the PC data relevant to the first orbit in the 0.3--10 keV 
band. A free constant factor was introduced to take into account both the
decrease of the mean flux between the WT and PC data and the different size of 
extraction regions. We obtain a photon index $\Gamma=1.77_{-0.10}^{+0.11}$ and a 
column density $N_{\rm H}=3.06_{-0.39}^{+0.43} \times 10^{21}$ cm$^{-2}$,
$\chi^2_{\rm red}=0.86$; 93 d.o.f. (Fig.~\ref{grb050721:xrtspec}). 
Figure~\ref{grb050721:xrtcont} shows the contour plot of the column density 
versus the power law photon index, which shows that the column density is 
higher ($>$ 3-$\sigma$ confidence level) than the Galactic value. The PC 
mode spectrum, obtained by the sum of the last 4 observations 
(see Tab~\ref{grb050721:obs}), yields a best fit value of 
$\Gamma=2.18_{-0.22}^{+0.25} $ and a column density consistent with the 
Galactic value ($\chi^2_{\rm red}=0.88$; 18 d.o.f.). The photon index 
of the late afterglow spectrum, compared with the early afterglow spectrum,
observed in the PC mode, shows a possible softening of the photon index 
by $\Delta\Gamma = 0.32\pm0.30$.


\section{Optical observations}

\begin{table*}[ht]
\centering
\caption{Log of optical observations.}
\begin{tabular}{ccccccccc} \hline
Run & Night       & UT start & UT end   & Filters	& Images     & Exp. Time [s] & Seeing     & Observing Mode \\ 
\hline
 1  & 2005 Jul 21 &  04:35:23 &  05:32:31 & $I_{\rm c}$      & 6                 & 120       & $<$ 1.4\arcsec & Robotic \\
 2  & 2005 Jul 21 & 04:53:58 & 05:45:21 & $R$		         & 34	               & 30          & 0.9\arcsec & VLT+RRM  \\
 3  & 2005 Jul 21 & 06:40:45 & 06:50:35 & $B$, $R$	         & 2, 4	      & 120, 60 & 1.2\arcsec & VLT+ToO  \\
 4  & 2005 Jul 22 & 00:47:11 & 04:39:39 & $B$, $R$, $I$ & 15, 20, 10   & 60, 10    & 0.6\arcsec & VLT+ToO  \\
 5  & 2005 Jul 23 & 02:00:42 & 02:07:41 & $R$		         & 5	               & 60	      & 1.5\arcsec & VLT+ToO  \\
 6  & 2005 Jul 24 & 00:31:57 & 01:51:58 & $R$, $I$	         & 10, 20         & 90	      & 0.7\arcsec & VLT+ToO  \\ 
 7  & 2005 Oct 06& 23:49:37 & 00:18:14 & $R$                   & 30               & 900        & 0.65\arcsec & VLT+ToO \\ 
 \hline
\end{tabular}
\label{tab_obs}
\end{table*}

\subsection{Prompt observation }

\begin{figure}
   \centering
   \includegraphics[width=9cm]{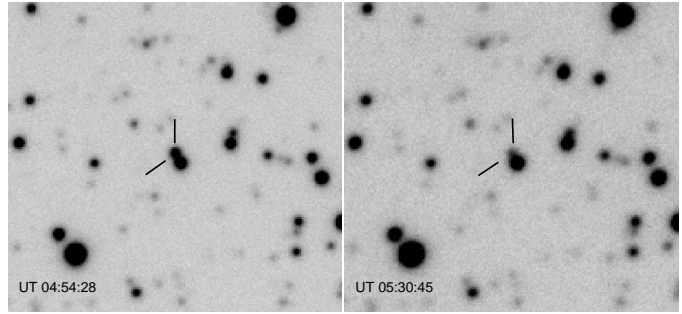}
\caption{VLT image of the field of the afterglow of GRB\,050721, showing 
                the proximity with the USNO star and the fading behavior of the 
                optical transient.}                
\label{fig:Afterglow}
\end{figure}

The remotely controlled 0.30m telescope, located in the New 
Mexico Skies observatory and operated from the Osaka University,
observed the field of GRB 050721 starting on 2005 July 21 at 
04:35:23 UT (369~s after the burst), collecting 6 $I_{\rm C}$ frames. 
A visual comparison of the frames with the DSS and 2MASS 
images revealed a bright, fading object within the XRT error circle.
Due to the large pixel scale (1\farcs41 per pixel, for a field 
of view of $24\arcmin \times 24\arcmin$) the optical afterglow was
indistinguishable from the USNO star (U0616--0429150) at
${\rm RA} = 16^{\rm h} 53^{\rm m} 44\fs496$, 
${\rm Dec} = -28\degr 22\arcmin 52\farcs75$ (J2000),
(heareafter star A, see Fig.~\ref{fig:Afterglow}). 
The contribution from star A was estimated from a second observation 
performed on 2005 July 25 with the same instrumental configuration and 
assuming a negligible contribution from the afterglow at that time. 

Here we report the first observation only in which the 
contribution of star A to the afterglow flux was negligible.
The observed OA flux was $I_c$=$15.6\pm0.2$ mag. 
The field was calibrated using USNO-B1.0 $I$ magnitudes and 
the total systematic error was estimated in absolute photometry as 
0.15~mag which is included in the error quoted above.

\subsection{Prompt and late VLT observations}

VLT optical data were obtained in five series during four nights, as
specified in Table~\ref{tab_obs}. The earliest data came from the 
activation of the rapid response mode, and consist of 33 images
in the $R$ filter, starting on 2005 Jul 21, 4:54 UT (25~min after the GRB) 
and lasting 53~min.
Since the RRM request also included a spectrum, we used the $R$-band
acquisition image for photometry too, hence reaching a total of 34 RRM
images in the $R$ filter. One hour later, a second series of 4 images in
$R$ and 2 in $B$ were acquired. In the two days after the burst,
target-of-opportunity (ToO) observations were activated to extend
the light-curve sampling. In total, the optical sample consists of 39
measurements in $R$, 2 in $I$, and 2 in $B$. 

\begin{table}[htbp]
\centering
\caption{Log of $R$ band observations.}
\begin{tabular}{cccc} 
\hline
Time since burst & Exp. time & Magnitude & Error   \\
             s                &       s         &       mag       &  mag   \\ 
\hline
1484 &  30  & 17.93 &  0.075\\
1559 &  30  & 18.00 &  0.075\\
1634 &  30  & 18.08 &  0.075\\
1710 &  30  & 18.17 &  0.074\\
1794 &  30  & 18.21 &  0.074\\
1869 &  30  & 18.26 &  0.075\\
1945 &  30  & 18.33 &  0.075\\
2021 &  30  & 18.39 &  0.073\\
2105 &  30  & 18.45 &  0.074\\
2181 &  30  & 18.50 &  0.074\\
2256 &  30  & 18.55 &  0.074\\
2332 &  30  & 18.62 &  0.074\\
2415 &  30  & 18.65 &  0.075\\
2491 &  30  & 18.70 &  0.074\\
2566 &  30  & 18.74 &  0.074\\
2642 &  30  & 18.81 &  0.075\\
2726 &  30  & 18.94 &  0.075\\
2802 &  30  & 18.86 &  0.074\\
2878 &  30  & 18.92 &  0.073\\
2954 &  30  & 18.95 &  0.074\\
3037 &  30  & 18.98 &  0.075\\
3113 &  30  & 19.01 &  0.075\\
3189 &  30  & 19.05 &  0.075\\
3265 &  30  & 19.03 &  0.075\\
3348 &  30  & 19.08 &  0.075\\
3424 &  30  & 19.12 &  0.077\\
3501 &  30  & 19.14 &  0.076\\
3577 &  30  & 19.20 &  0.075\\
3661 &  30  & 19.18 &  0.074\\
3737 &  30  & 19.19 &  0.074\\
3813 &  30  & 19.29 &  0.075\\
3889 &  30  & 19.35 &  0.077\\
3973 &  30  & 19.44 &  0.077\\
4537 &  30  & 19.48 &  0.043\\
7891 &  240 & 20.28 &  0.077\\
73076&  1000& 21.58 &  0.016\\
78075&  20  & 21.66 &  0.062\\
85946&  1000& 21.60 &  0.024\\
248595& 1200& 22.51 &  0.110\\
6721500& 900 & $>$25.8 & (5$\sigma$ u.l.)\\
\hline
\end{tabular}
\label{tab_obs}
\end{table}

\begin{figure}[ht!]
   \centering
   \includegraphics[width=9.cm, angle=0]{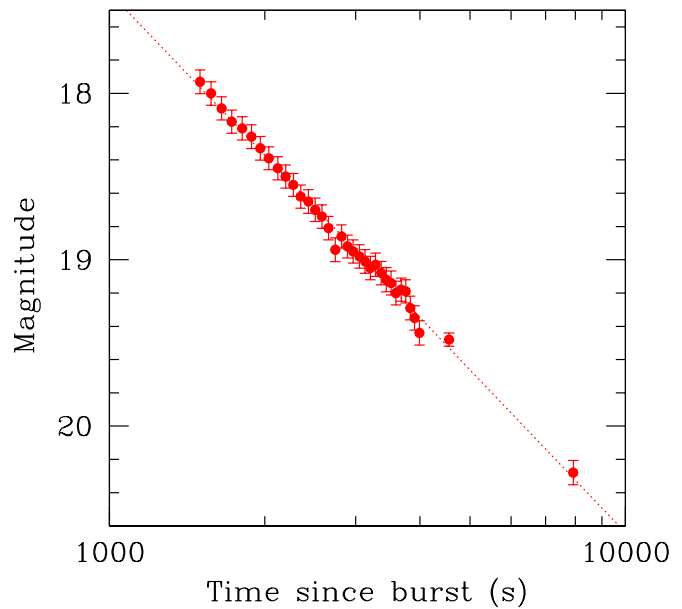}
   \caption{The $R$-band light curve of the early afterglow of GRB\,050721 
   observed with VLT equipped with FORS\,2. Optical data can be fitted by 
   a simple power law model ($F(t)\propto t^{-\alpha}$) with $\alpha = 1.29 \pm 0.06$
   (dotted line).
   This slope is in good agreement with the X-ray slope at the same time.}
   \label{fig:earlyag}
\end{figure}

\subsubsection{Optical data reduction and analysis}

The optical afterglow was detected at 
${\rm RA} $=$16^{\rm h} 53^{\rm m} 44\farcs53$, 
${\rm Dec}$ =$-28^{\circ} 22\arcmin 51\farcs8$ (J2000),
very close (1\farcs4) to a relatively bright star ($R$=17.4) present 
in the USNO\,B1.0 catalog (U0616--0429150), see Fig. \ref{fig:Afterglow}).  
In order to avoid any possible contribution from the nearby star, a PSF 
profile fitting was adopted using \texttt{DAOPHOT/ALLSTAR} 
(\cite{stetson87, stetson94}) within IRAF\footnote{IRAF is distributed by 
the National Optical Astronomy Observatories, which are operated by 
the Association of Universities for Research in Astronomy, Inc., under 
cooperative agreement with the National Science Foundation}. In order 
to maximize the efficiency in the detection, all the $R$-band images 
from the RRM run were carefully aligned and averaged together. 
Since the FWHM varied slightly during the acquisition of the
sequence, only the best-seeing frames were used to obtain the average
master image. The master was then used to create, with \texttt{daofind},
a list of candidate objects, that was used as input list to process the
individual images.  Thanks to the good mean seeing conditions,
\texttt{DAOPHOT/ALLSTAR} was able to resolve the two
components of the pair USNO star/afterglow, providing an excellent
PSF fitting. The output photometry lists were then carefully checked
for relative zeropoint differences, finding a very good agreement
among the measurements without the need of further re-adjustment. All the
catalogs from the individual images were translated to a homogeneous
coordinate system by transforming pixel coordinates into sky
coordinates. The same approach was also used later for the
images obtained in the other ToO series. Magnitudes and, where
applicable, colors have been obtained for all the objects in the
master candidate list. FORS\,1 and FORS\,2 data have been later matched by
transforming all the image coordinates to sky coordinates, obtaining a
homogeneous catalog. Image calibration was obtained after applying
aperture corrections to the measured objects; calibration relations derived
from standard star observations were applied. Standard stars were observed
all nights and the calibrations were performed with the IRAF package
\texttt{photcal}. Color terms have been fitted against the available magnitudes,
i.e. $B-R$ for the $B$ and $R$ bands, and $R-I$ for the $I$ band, and were
found to be very small (+0.015 for $B-R$, $< 0.005$ for $R-I$). Since the
standards (Landolt fields PG\,1323-086 and SA\,110) were observed only
once per night, we used the mean extinction coefficients for Paranal.

\begin{figure}[ht]
   \centering
   \includegraphics[width=9cm]{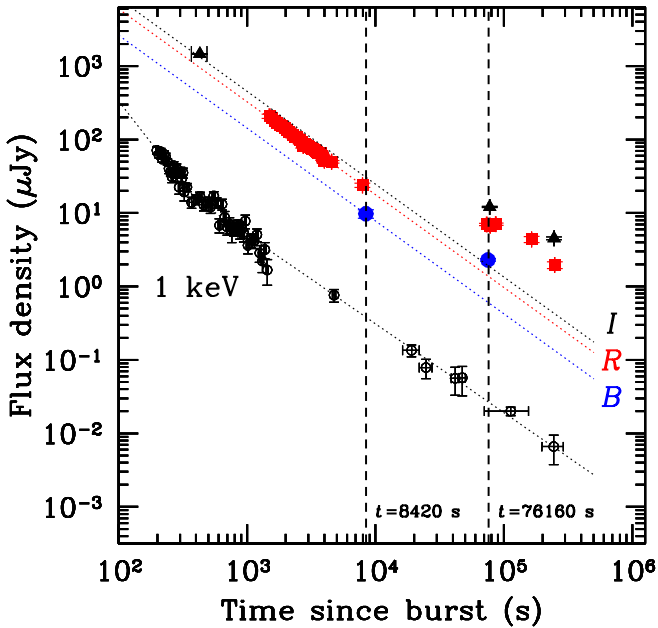}
   \caption{Multiwavelength light curve of the afterglow of GRB 050721: from top to bottom 
   $I$ band (black filled triangles), $R$ band (red filled squares), $B$ band (blue filled circles),
   X-flux at 1 keV (black open circles). Vertical dashed lines show the times of the SEDs in Fig. \ref{fig:sed}.}
   \label{fig:lc}
\end{figure}

The VLT prompt observations, obtained in RRM, provided a very well sampled light
curve. The observed light curve shows the typical fading behavior for GRB afterglows 
well described by a simple power law  ($F(t) = K t^{-\alpha}$)  with a slope $\alpha = 1.29 \pm 0.06$
which is in good agreement with the X-ray slope during the same time interval (see Fig. \ref{fig:lc}). 
A late point, obtained at the end of the night ($\sim 8000$~s after the burst and 
$\sim 6500$~s after the first point), is well aligned along the early curve. The optical light 
curve is smoothly decaying and it does not show any significant flaring activity. GRB\,050721 was 
also observed during the second night starting about 20~hr after the burst. We collected our data 
set in three different images in which the afterglow is still clearly detected and we found it 
about $1.8$~mag brighter than the value predicted by extrapolating the first night data set. 
Such a rebrightening was also observed in the following two observations, obtained 
respectively 45 and 69~hr after the burst. Moreover if we consider these
two points only we obtain a power law decay with a steeper slope ($\alpha = 1.9 \pm 0.7$)
but still consistent with the previous one given the large error. A late deep observation 
(2.5 months later) did not reveal either an afterglow or a host galaxy down to a limiting 
magnitude of $R > 25.8$~mag (5$\sigma$). We can conclude that the observed flattening  
was not due to the emergence of the host galaxy, but that a rebrightening was present in 
the afterglow.

\begin{figure}[ht]
   \centering
   \includegraphics[width=9cm]{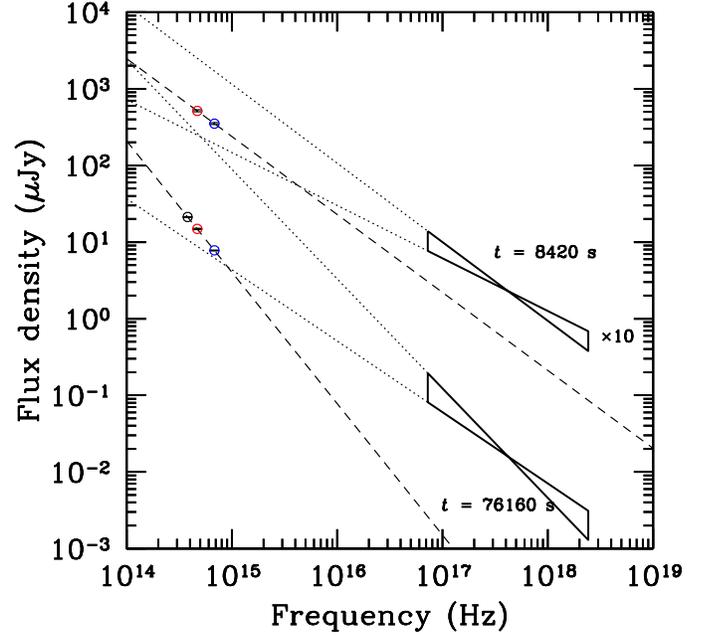}
   \caption{Spectral energy distribution of the afterglow of GRB 050721 before the optical bump
   ($t=8420$ s)  and at the time of the optical bump ($t=76\,160$ s).  Early optical and X-ray fluxes 
   are artificially shifted up by a factor of 10 to distance them from the later time points.
      }
   \label{fig:sed}
\end{figure}

\section{Results and discussion}

GRB\,050721 was promptly observed in both the X-ray and optical bands and 
it was accurately monitored for several days after the burst. Such a long term 
multiwavelength monitoring showed a different behavior in the X-ray and
optical light curves. Thanks to Swift, it is now well known that 
the early X-ray light curves of a large fraction of the bursts are characterized by a 
steep decay, followed by a shallower phase and finally a somewhat steeper 
decay (\cite{taglia05, Nousek05, OBrien05}). Many events have superimposed 
X-ray flares, probably still related to  the central engine (\cite{Burrows05, Falcone06, 
romano06, Zhang05, guetta06}). In the case of GRB\,050721 we did not detect flares in the X-ray
light curve, while we saw the early steep decay, although not one of the steepest
so far seen by XRT, followed by a shallower phase at about 400 s after the burst.
We also did not see a second break in the 0.2--10.0 keV light curve, which in fact
shows a regular fading behaviour up to a time interval of about one week, that can 
be fitted with a simple power law ($F(t) = K t^{-\alpha}$) yielding a slope 
of $\alpha=1.20\pm0.04$ on the entire observation. The optical light curve shows 
the same regular fading behavior over a time interval of about 7000~s with the 
same power law ($\alpha = 1.29 \pm 0.06$) then, at about 24 hours after the burst,
a rebrightening of about 1.8 mag is observed.

The consistency between the early optical and X-ray decay indices suggests that both emissions
result from the same component during the first hours after the GRB. 
This fact can be futher confirmed by looking at the broadband spectral energy distribution (SED). 
Figure~\ref{fig:sed} shows
the SED at two times, chosen during the initial decay phase and during the optical  bump. At
the first epoch, the optical data agree well with the extrapolation of the X-ray spectrum. The
optical color ($B-R = 1.25$~mag) also provides a spectral index $\beta_{\rm opt} =
1.16 \pm 0.35$, after correcting for the significant Galactic extinction ($A_B = 1.2$~mag,
$A_R = 0.75$~mag; Schlegel et al. 1998). 
This value, given its large uncertainty, is consistent with that observed in the X-ray range 
($\beta_{\rm X} = 0.86 \pm 0.17$), and with the broad-band optical-to-X-ray slope 
($\beta_{\rm OX} = 0.80$). The picture is therefore consistent with the optical and X-ray 
emission lying on the same segment of a power-law SED, with slope $\beta = 0.80$. 
Thus, no synchrotron breaks were present between the two bands. Moreover, the 
extinction affecting the optical data was likely small.

The spectral and temporal properties of the afterglow of GRB\,050721 during the first hours
are consistent with a fireball expanding inside a uniform external medium (\cite{Sari98}), 
providing that the cooling frequency lies blueward of the X-ray band. 
For a spectral index $\beta \sim 0.80$, the predicted decay slope is $\alpha = 3\beta/2 \sim 1.2$, 
in very good agreement with the observed values both in the optical and in the X-ray band (their 
average being $\alpha = 1.23 \pm 0.03$). The inferred electron distribution index is computed as 
$p = 2\beta + 1 \sim 2.6$, or, alternatively, $p = 4\alpha/3 + 1 = 2.64 \pm 0.04$. Such value is not 
uncommon among GRB afterglows.

The situation changes substantially starting $\sim 1$~d after the GRB. At this time, the
optical flux is significantly brighter than predicted by the early-time decay and the optical
spectrum is softer ($\beta_{\rm opt} = 1.85 \pm 0.11$) than observed at earlier time, and steeper 
than usually observed in GRB afterglows. Both the flux and SED suggest that a new 
mechanism is powering the optical emission, leaving unaffected the X-ray region. 

Among the possible explanations for a bump in the optical light curve,
we consider the emergence of an additional energy source like a
supernova (SN). It is now well established that long duration GRBs
(or a significant fraction of them) are produced in SN explosions
(\cite{Galama98,Stanek03,Hjorth03,Malesani04}).
Sometime bumps have been detected during the afterglow
decay days/weeks past the GRB (e.g. \cite{Bloom99,MdV06}).
However, the bump associated with GRB 050721 rose very
quickly, only $\sim 1$~d after the gamma event, which is much shorter
than the observed rising time of SNe-Ibc (10-20 days, e.g. \cite{hamuy03}).  
One possibility is that the SN has occurred before the GRB, as
advocated in the Supranova model (\cite{VietriStella99}). An analysis
of the present data suggest that SNe and GRBs go off simultaneously
(\cite{mdv05}), nevertheless a few days of delay between SN and GRB 
events can be yet accommodated within the uncertainties. The major
caveat for this hypothesis is represented by the magnitude of the bump 
($R\approx21$). For a SN as bright as SN\,1998bw ($M_R = -19.4$;
\cite{Galama98}), the inferred redshift would be $z \sim 0.25$. This would be difficult to
reconcile with the nondetection of any host galaxy (our limit would correspond to $0.003L_*$
at this redshift). A fainter SN would imply a distance even smaller. We thus consider this
hypothesis unlikely.

A more exotic possibility calls for the mini-SN scenario (\cite{LiPac98}), 
which evolutionary timescale is comparable to that of
the rebrightening (1--2 d). However this scenario would suffer a problem 
similar to the one faced by a conventional SN, since the peak magnitude is 
expected to be comparable to that of a SN.

A plausible explanation for the rebrightening could be given by modifications 
in the afterglow physics: rebrightenings due to energy injection (e.g. refreshed shock,  
\cite{guetta06}) or fluctuations in the external medium. In the refreshed shock scenario
flux fluctuations in both the X-ray and optical bands are expected to be proportional to 
energy fluctuations of the fireball. Therefore, a bump in both the optical and X-ray light curves
is expected. The lack of a rebrightening in the X-ray band rules out this hypothesis. 

In the case of fluctuations in the external medium, we expect a bump in the optical 
light curve (but not in the X-ray one) when the fireball encounters regions with 
enhanced density $n$ (e.g. \cite{Lazzati02, Berg00}).  Such a density jumps may
be caused by winds termination shock, as proposed for the case of GRB 030329 
(\cite{huang06}). Nakar, Granot \& Piran (2003) have 
shown that the afterglow flux $F_{\nu}$ at different frequencies may be written in 
terms of the external density in a very simple way. Assuming the typical afterglow 
parameters,  the synchrotron cooling frequency at $1$ d from the burst is expected 
to be $\nu_{\rm c}\sim 10^{15}$~Hz (just between optical and X-ray band). 
Our data on the late X-ray afterglows of GRB\,050721 cannot exclude (within 
uncertainties) both a steepening in the light curve ($\alpha=1.20^{+0.39}_{-0.42}$)
and a softening of the photon index ($\Delta\Gamma = 0.32\pm0.30$), that are 
compatible with the hypothesis that $\nu_{\rm c}$ is shifted to energies lower than the 
X-ray band at this epoch. In such a case the 
X-ray flux ($\nu > \nu_{\rm c}$) is weakly dependent on the external medium 
density $n$, while the optical flux ($\nu < \nu_{\rm c}$ ) is  $F_\nu \propto n^{3/4}$ 
(\cite{Nakar03}). Since we see a rebrightening only in the optical band 
($\nu < \nu_{\rm c}$) and  not in the X-ray band ($\nu > \nu_{\rm c}$) we 
can conclude that this is an evidence of fluctuations in the external medium. 
In particular we can derive an estimate of the variation in the density from the 
fluctuations in the flux $ n_2/n_1 \propto (F_{\nu}^{(2)}/F_{\nu}^{(1)})^{4/3}\sim 10$.

However, the optical flux at the second epoch shows a redder spectral distribution than 
what expected by adopting a simple synchrotron model with parameters extrapolated 
by the X-ray flux. A possible explanation for such a red spectrum is that the high-density 
region could be composed of very dense clumps, with a covering factor $f \ll 1$. In such 
a scenario, the rebrightening observed in the optical band may be due to the interaction 
of the fireball with the clumps. Dust in the clumps may survive the intense radiation field, 
so that the spectrum produced inside them emerges reddened. Since $f \ll 1$the optical 
flux did not suffer significant extinction, since only a fraction $f$ was intercepted by the 
high-density clouds.

\section{Conclusions}

We have presented a comprehensive multiwavelength study 
of the afterglow of GRB 050721 based on RRM and ToO 
observation with the VLT, early robotic observations and 
Swift XRT data. Both the X-ray and optical light-curve 
are very well sampled and showed a regular fading behavior
(with almost the same decay index) within 24 hrs after 
the burst.  At about one day from the burst a remarkable 
(1.8 mag) rebrightening is observed in the optical afterglow,
while no variations are observed in the X-ray light curve at
the same epoch. A late optical observation (2.5 months after the
burst) did not show any afterglow or host galaxy down 
to a limiting magnitude of $R >$25.8 (5$\sigma$) confirming 
that the observed rebrightening was not due to  the presence of a
bright host galaxy. The broad band SED obtained at two different 
epochs, during the initial decay and during the rebrightening, suggests 
that during the second epoch a new mechanism is powering the
optical emission, leaving unaffected the X-ray region. Many 
different scenarios are considered in order to explain such a 
peculiar rebrightening.  Since the rebrightening was observed 
only in the optical band and not in the X-ray band we propose 
that this is a plausible evidence of fluctuations in the external 
medium and we derive an estimate of the variation in the 
density from the fluctuations in the flux $ n_2/n_1 \sim 10$. 
In order to take into account  an optical spectral distribution 
redder than what expected by adopting a simple synchrotron 
model, we suggest that the observed rebrightening may be due 
to the interaction of the fireball with a high-density region 
composed by very dense clumps, with a covering factor $f \ll 1$.

\begin{acknowledgements}

This work is supported at OAR, OAB and IFC-PA by funding from ASI on grant 
number I/R/039/04, at Penn State by NASA contract NAS5-00136 and at the 
University of Leicester by the PPARC on grant numbers PPA/G/S/00524
and PPA/Z/S/2003/00507. This research was supported in part by the National 
Science Foundation under Grant No. PHY99-0794. We also thank Alex Kann and 
an anonymous referee for their helpful comments. We gratefully acknowledge the 
support from the Swift team and from the ESO VLT staff. 

\end{acknowledgements}


\begin{thebibliography} {\protect\astroncite{Stetson}{1987}}

\bibitem[Antonelli et al. 2005]{anto05} Antonelli, L.A., Page, K., Morris, D., et al. 2005, GCN 3654

\bibitem[Barthelmy et al. 2005]{barthelmy05} Barthelmy, S., Barbier, L.M., Cummings, J.R., et al. 2005, \ssr, 120, 143

\bibitem[Beloborodov 2003]{Belo03} Beloborodov, A.M. 2003, \apj, 585, L19

\bibitem[Berger et al.,  2000]{Berg00} Berger, E. et al., 2000, \apj, 545, 56

\bibitem[Bloom et al. 1999]{Bloom99} Bloom, J.S., Kulkarni, S.R., Djorgovski, S.G., et al. 1999, \nat, 401, 453

\bibitem[Burrows et al. 2005]{burr05} Burrows, D.N., Hill, J.E., Nousek, J.A., et al. 2005, \ssr, 120, 165

\bibitem[Burrows et al. 2005]{Burrows05} Burrows, D.N., Hill, J.E., Chincarini, G., et al. 2005, \apj, 622, L85

\bibitem[Covino et al. 2005]{covi05} Covino, S., D'Avanzo, P., Bagnulo, S., et al. 2005, GCN 3656

\bibitem[D'Avanzo et al. 2005]{davi05} D'Avanzo, P., Covino, S., Malesani, D., et al. 2005, GCN 3658

\bibitem[Della Valle et al. 2005]{mdv05} Della Valle, M., 2005, in the Proceedings of the 4th Workshop Gamma-Ray Bursts in the Afterglow Era, Rome, 18-22 October 2004. Editors: L. Piro, L. Amati, S. Covino, and B.  Gendre. Il Nuovo Cimento,  28, 563

\bibitem[Della Valle et al. 2006]{MdV06} Della Valle, M., Malesani, D., Bloom, J.S., et al. 2006, \apjl, submitted

\bibitem[Dickey \& Lockman 1990]{dicklock90} Dickey, J.M., \& Lockman, F.J. 1990, \araa, 28, 215

\bibitem[Fan et al. 2005]{Fan05} Fan, Y.Z., Zhang, B., \& Wei, D.M. 2005, \apj, 628, 298

\bibitem[Falcone et al. 2006]{Falcone06} Falcone, A., Burrows, D.N., Lazzati, D., et al. 2006, \apj, in press (astro-ph/0512615)

\bibitem[Fenimore et al. 2005]{feni05} Fenimore, E., Barbier, L.,  Barthelmy, S., et al. 2005, GCN 3661

\bibitem[Fox et al. 2003]{Fox03} Fox, D., Yost, S., Kulkarni, S.R., et al. 2003, \nat, 422, 284

\bibitem[Galama et al. 1998]{Galama98} Galama, T.J., Vreeswijk, P.M., van Paradijs, J., et al. 1998, \nat, 395, 670

\bibitem[Gehrels et al. 2004]{gehr04} Gehrels, N., Chincarini, G., Giommi, P., et al. 2004, \apj, 611, 1005

\bibitem[Guetta et al.  2006]{guetta06} Guetta, D., Fiore, F., D'Elia, V., et al., 2006, \aap, submitted (astro-ph/0602387)

\bibitem[Hamuy,  2003]{hamuy03} Hamuy, M. 2003, in the Proceedings of ÕÕCore Collapse of Massive Stars'', Ed. C.L. Fryer, Kluwer, Dordrecht (astro-ph/0301006)

\bibitem[Henden 2000]{henden00} Henden, A. 2000, Journ. AVSO, 29, 35

\bibitem[Hjorth et al. 2003]{Hjorth03} Hjorth, J., Sollerman, J., M{\o}ller, P., et al. 2003, \nat, 423, 847

\bibitem[Huang et al. 2006]{huang06} Huang, Y.F., Cheng, K.S., \& Gao, T.T., 2006, \apj, 637, 873.

\bibitem[Kalberla et al. 2005]{kalberla05} Kalberla, P.M.W., Burton, W.B., Hartmann, D., et al. 2005, \aap, 440, 775

\bibitem[Lazzati et al. 2002]{Lazzati02} Lazzati, D., Rossi, E., Covino, S., Ghisellini, G., \& Malesani, D. 2002, \aap, 396, L5

\bibitem[Li \& Paczy\'nski 1998]{LiPac98} Li, L.-X., \& Paczy\'nski, B. 1998, \apj, 507, L59

\bibitem[Malesani et al. 2004]{Malesani04} Malesani, D., Tagliaferri, G., Chincarini, G., et al. 2004, \apj, 609, L5

\bibitem[Mazzali et al. 2002]{Mazzali02} Mazzali, P.A., Deng, J., Maeda, K., et al. 2002, \apj, 572, L61

\bibitem[Moran \& Reichart 2005]{Moran05} Moran, J.A., \& Reichart, D.E. 2006, \apj, 632, 438

\bibitem[Moretti et al. 2005]{centroids}  Moretti, A., Perri, M., Capalbi, M., et al.\ 2005, \aap, 448, L9

\bibitem[Nakar et al. 2003]{Nakar03} Nakar, E., Piran, T., \& Granot, J. 2003, New Astr., 8, 495

\bibitem[Nousek et al. 2006]{Nousek05} Nousek, J.A., Kouveliotou, C., Grupe, D., et al. 2006, \apj, in press (astro-ph/0508332)

\bibitem[O'Brien et al. 2006]{OBrien05} O'Brien, P.T., Willingale, R., Osborne, J., et al. 2006, \apj, submitted (astro-ph/0512615)

\bibitem[Romano et al. 2005]{pat05} Romano, P., Antonelli, L.A., Chincarini, G., et al. 2005, GCN 3659

\bibitem[Romano et al. 2006]{romano06} Romano, P.,  et al. 2006, \apj, in press (astro-ph/0601773)

\bibitem[Roming et al. 2005]{roming05} Roming, P.W.A., Kennedy, T.E., Mason, K.O., et al. 2005, \ssr, 120, 95

\bibitem[Sari, Piran, \& Narayan, 1998]{Sari98} Sari, R., Piran, T., \& Narayan, R. 1998, \apj, 497, L17

\bibitem[Sari \& M\'esz\'aros 2000]{SariMeszaros00} Sari, R., \& M\'esz\'aros 2000, \apj, 535, L33

\bibitem[Schegel et al. 1998]{Schlegel98} Schlegel, D.J., Finkbeiner, D.P., \& Davies, M. 1998, \apj, 500, 525

\bibitem[Stanek et al. 2003]{Stanek03} Stanek, K.Z., Matheson, T., Garnavich, P.M., et al. 2003, \apj, 591, L17

\bibitem[Stetson 1987]{stetson87} Stetson, P.B.E. 1987, \pasp, 99, 191

\bibitem[Stetson 1994]{stetson94} Stetson, P.B., 1994, \pasp, 106, 250

\bibitem[Tagliaferri et al. 2005]{taglia05} Tagliaferri, G., Goad, M., Chincarini, G., et al., 2005, \nat, 436, 985

\bibitem[Testa et al. 2005]{test05} Testa, V., Antonelli, L.A., Malesani, D., et al. 2005, GCN 3662

\bibitem[Torii 2005]{tori05} Torii, K. 2005, GCN 3655

\bibitem[Vietri \& Stella 1999]{VietriStella99} Vietri, M., \& Stella, L. 1999, \apj, 527, L43

\bibitem[Waxman \& Draine 2000]{WaxmanDraine00} Waxman, E., \& Draine, B.T. 2000, \apj, 537, 796

\bibitem[Zhang et al. 2006]{Zhang05} Zhang, B., Fan, Y., Dyks, J., et al. 2006, \apj, in press (astro-ph/0508321)

\end{thebibliography}
\end{document}